\documentclass[journal]{IEEEtran}

\usepackage[dvipsnames]{xcolor}
\usepackage{siunitx}

\usepackage{cite}

\usepackage{graphicx}
\usepackage[cmex10]{amsmath}
\usepackage{amssymb,mathabx}
\usepackage{bbm}

\usepackage{array}
\usepackage{booktabs}
\usepackage[caption=false,font=normalsize,labelfont=sf,textfont=sf]{subfig}

\usepackage[ddmmyyyy]{datetime}
\newcommand{\header}{PREPRINT, \today, \currenttime}

\DeclareMathOperator*{\argmax}{\mathrm{argmax}}

\begin{document}
\title{End-to-end Deep Learning of\\ Optical Fiber Communications}

\author{Boris~Karanov,~\IEEEmembership{Student Member,~IEEE,}
        Mathieu~Chagnon,~\IEEEmembership{Member,~IEEE,}
        F{\'e}lix~Thouin,~\IEEEmembership{}
        Tobias~A.~Eriksson,~\IEEEmembership{}
        Henning~B{\"u}low,~\IEEEmembership{}
        Domani\c{c}~Lavery,~\IEEEmembership{Member,~IEEE,}
        Polina~Bayvel,~\IEEEmembership{Fellow,~IEEE,}
        and~Laurent~Schmalen,~\IEEEmembership{Senior~Member,~IEEE}
\thanks{B.~Karanov is with Nokia Bell Labs, Stuttgart, Germany and Optical Networks Group (ONG), Dept. Electronic and Electrical Engineering, University College London (UCL), UK (email:boris.karanov.16@ucl.ac.uk).
M.~Chagnon, H.~B{\"u}low and L.~Schmalen are with Nokia Bell Labs, Stuttgart, Germany.
F.~Thouin is with the School of Physics, Georgia Institute of Technology, Atlanta, GA, USA.
T.~A.~Eriksson is with the Quantum ICT Advanced Development Center, National Institute of Information and Communications Technology (NICT), 4-2-1 Nukui-kita, Koganei, Tokyo 184-8795, Japan.
D.~Lavery and P.~Bayvel are with ONG, Dept. Electronic and Electrical Engineering, UCL, UK.}}

\markboth{\header}{\header}

\maketitle

\begin{abstract}
In this paper, we implement an optical fiber communication system as an end-to-end deep neural network, including the complete chain of transmitter, channel model, and receiver. This approach enables the optimization of the transceiver in  a single end-to-end process. We illustrate the benefits of this method by applying it to intensity modulation/direct detection (IM/DD) systems and show that we can achieve bit error rates below the 6.7\% hard-decision forward error correction (HD-FEC) threshold. We model all componentry of the transmitter and receiver, as well as the fiber channel, and apply deep learning to find transmitter and receiver configurations minimizing the symbol error rate. We propose and verify in simulations a training method that yields robust and flexible transceivers that allow---without reconfiguration---reliable transmission over a large range of link dispersions. The results from end-to-end deep learning are successfully verified for the first time in an experiment. In particular, we achieve information rates of 42\,Gb/s below the HD-FEC threshold at distances beyond 40\,km. We find that our results  outperform conventional IM/DD solutions based on 2 and 4 level pulse amplitude modulation (PAM2/PAM4) with feedforward equalization (FFE) at the receiver.  Our study is the first step towards end-to-end deep learning-based optimization of optical fiber communication systems.
\end{abstract}


\IEEEpeerreviewmaketitle

\section{Introduction}

\IEEEPARstart{T}{he} application of machine learning techniques in communication systems has attracted a lot of attention in recent years \cite{Jiang,Khan2}. In the field of optical fiber communications, various tasks such as performance monitoring, fiber nonlinearity mitigation, carrier recovery and modulation format recognition have been addressed from the machine learning perspective \cite{Thrane,Zibar,Khan}. In particular, since chromatic dispersion and nonlinear Kerr effects in the fiber are regarded as the major information rate-limiting factors in modern optical communication systems~\cite{Essiambre}, the application of artificial neural networks~(ANNs), known as universal function approximators~\cite{Hornik}, for channel equalization has been of great research interest~\cite{Burse,Ibnkahla,Jarajreh,Giacoumidis,Eriksson}. For example, a multi-layer ANN architecture, which enables deep learning techniques~\cite{Goodfellow}, has been recently considered in \cite{Hager} for the realization of low-complexity nonlinearity compensation by digital backpropagation~(DBP)~\cite{Ip}. It has been shown that the proposed ANN-based DBP achieves similar performance than conventional DBP for a single channel 16-QAM system while reducing the computational demands. Deep learning has also been considered for short-reach communications. For instance, in~\cite{Gaiarin} ANNs are considered for equalization in PAM8 IM/DD systems. Bit-error rates (BERs) below the forward error correction (FEC) threshold have been experimentally demonstrated over 4\,km transmission distance. In~\cite{Estaran}, deep ANNs are used at the receiver of the IM/DD system as an advanced detection block, which accounts for channel memory and linear and nonlinear signal distortions. For short reaches (1.5\,km), BER improvements over common feed-forward linear equalization were achieved.   

In all the aforementioned examples, deep learning techniques have been applied to optimize a specific function in the fiber-optic system, which itself consists of several signal processing blocks at both transmitter and receiver, each carrying out an individual task, e.g. coding, modulation and equalization. In principle, such a modular implementation allows the system components to be analyzed, optimized and controlled separately and thus presents a convenient way of building the communication link. Nevertheless, this approach can be sub-optimal, especially for communication systems where the optimum receivers or optimum blocks are not known or not available due to complexity reasons. As a consequence, in some systems, a block-based receiver with one or several sub-optimum modules does not necessarily achieve the optimal end-to-end system performance. Especially if the optimum joint receiver is not known or too complex to implement, we require carefully chosen approximations.

Deep learning techniques, which can approximate any nonlinear function~\cite{Goodfellow}, allow us to design the communication system by carrying out the optimization in a single end-to-end process including the transmitter and receiver as well as the communication channel. Such a novel design based on full system learning avoids the conventional modular structure, because the system is implemented as a single deep neural network, and has the potential to achieve an optimal end-to-end performance. The objective of this approach is to acquire a robust representation of the input message at every layer of the network. Importantly, this enables a communication system to be adapted for information transmission over any type of channel without requiring prior mathematical modeling and analysis. The viability of such an approach has been introduced for wireless communications~\cite{O'Shea} and also demonstrated experimentally with a wireless link~\cite{Doerner}. Such an application of end-to-end deep learning presents the opportunity to fundamentally reconsider optical communication system design.

Our work introduces end-to-end deep learning for designing optical fiber communication transceivers. The focus in this paper is on IM/DD systems, which are currently the preferred choice in many data center, access, metro and backhaul applications because of their simplicity and cost-effectiveness~\cite{Eiselt}. The IM/DD communication channel is nonlinear due to the combination of photodiode (square-law) detection and fiber dispersion. Moreover, noise is added by the amplifier and the quantization in both the digital-to-analog converters (DACs) and analog-to-digital converters (ADCs). We model the fiber-optic system as a deep fully-connected feedforward ANN. Our work shows that such a deep learning system including transmitter, receiver, and the nonlinear channel, achieves reliable communication below FEC thresholds. We experimentally demonstrate the feasibility of the approach and achieve information rates of 42\,Gb/s beyond 40\,km. We apply re-training of the receiver to account for the specific characteristics of the experimental setup not covered by the model.  Moreover, we present a training method for realizing flexible and robust transceivers that work over a range of distance. Precise waveform generation is an important aspect in such an end-to-end system design. In contrast to~\cite{O'Shea}, we do not generate modulation symbols, but perform a direct mapping of the input messages to a set of robust transmit waveforms. 

{The goal of this paper is to design, in an offline process, transceivers for low-cost optical communication system that can be deployed without requiring the implementation of a training process in the final product. During the offline training process, we can label the set of data used for finding the parameters of the ANN and hence use \emph{supervised} training. This is a first step towards building a deep learning-based optical communication system. Such a system will be optimized for a specific range of operating conditions. Eventually, in future work, an online training may be incorporated into the transceiver, which may still work in a supervised manner using, e.g., pilot sequences, to cover a wider range of operating conditions. Building a complete \emph{unsupervised} transceiver with online training will be a significantly more challenging task and first requires a thorough understanding of the possibilities with supervised training. Hence, we focus on the supervised, offline training case in this paper.}

The rest of the manuscript is structured as follows: Section~\ref{sec:ANN} introduces the main concepts behind the deep learning techniques used in this work. The IM/DD communication channel and system components are described mathematically in Sec.~\ref{sec:system}. The architecture of the proposed ANN along with the training method is also presented in this section. Section~\ref{sec:simulation} reports the system performance results in simulation. Section\,\ref{sec:experiment} presents the experimental test-bed and validation of the key simulation results. {Section~\ref{sec:discussion} contains an extensive discussion on the properties of the transmit signal, the advantages of training the system in an end-to-end manner, and the details about the experimental validation.} Finally, Sec.~\ref{sec:conclusion} concludes the work.

\section{Deep Fully-connected Feed-forward Artificial Neural Networks}\label{sec:ANN}
A fully-connected $K$-layer feed-forward ANN maps an input vector $\mathbf{s}_{0}$ to an output vector $\mathbf{s}_{K}=f_{\textrm{ANN}}(\mathbf{s}_{0})$ through iterative steps of the form
\begin{equation}
\label{eq:LayerMapping}
\mathbf{s}_{k}=\alpha_{k}(\mathbf{W}_{k} \mathbf{s}_{k-1}+\mathbf{b}_k),\qquad k=1,..,K.
\end{equation}
Where $\mathbf{s}_{k-1}\in{\mathbb{R}}^{N_{k-1}}$ is the output of the $(k-1)$-th layer, $\mathbf{s}_{k}\in{\mathbb{R}}^{N_k}$ is the output of the $k$-th layer, $\mathbf{W}_{k}\in{\mathbb{R}}^{N_{k}\times N_{k-1}}$ and $\mathbf{b}_{k}\in{\mathbb{R}}^{N_k}$ are respectively the weight matrix and the bias vector of the $k$-th layer and $\alpha_{k}$ is its activation function. The set of layer parameters $\mathbf{W}_k$ and $\mathbf{b}_k$ is denoted by
\begin{equation}
\label{eq:LayerParams}
\boldsymbol{\theta}_{k} = \{\mathbf{W}_k,\mathbf{b}_k\}.
\end{equation}
The activation function $\alpha_{k}$ introduces nonlinear relations between the layers and enables the approximation of nonlinear functions by the network. A commonly chosen activation function in state-of-the-art ANNs is the rectified linear unit (ReLU), which acts individually on each of its input vector elements by keeping the positive values and equating the negative to zero~\cite{Nair}, i.e., $\mathbf{y}=\alpha_{\tiny{\text{ReLU}}}(\mathbf{x})$ with
\begin{equation}
\label{eq:ReLuActivation}
y_{i}=\max(0,x_i),
\end{equation}
where $y_i$, $x_i$ denote the $i$-th elements of the vectors $\mathbf{y}$ and $\mathbf{x}$, respectively. Compared to other popular activation functions such as the hyperbolic tangent and sigmoid, the ReLU function has a constant gradient, which renders training computationally less expensive and avoids the effect of \emph{vanishing gradients}. {This effect occurs for activation functions with asymptotic behavior since the gradient can become small and consequently decelerate the convergence of the learning algorithm~\cite[Sec. 8.2]{Goodfellow}.} 

\begin{figure*}[ht]
	\centering
	\includegraphics[width=\textwidth]{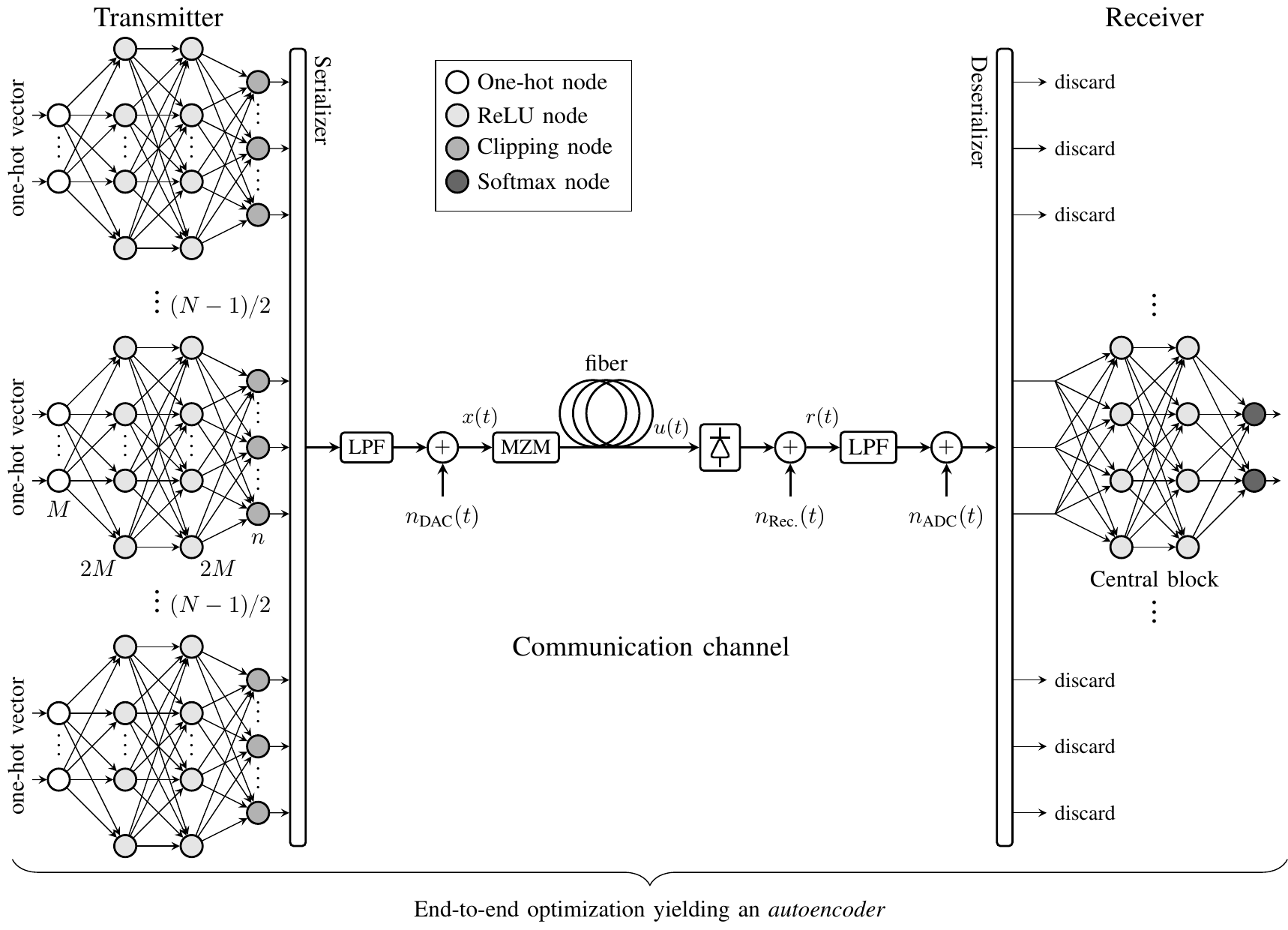}
      \caption{Schematic of the IM/DD optical fiber communication system implemented as a deep fully-connected feedforward neural network. Optimization is performed between the input messages and the outputs of the receiver, thus enabling end-to-end deep learning of the complete system.}
       \label{fig:Schematic}
\end{figure*}
The final (decision) layer of an ANN often uses the \emph{softmax} activation function, where the elements $y_i$ of the output $\mathbf{y}=\mathop{\textrm{softmax}}(\mathbf{x})$ are given by
\begin{equation}
\label{eq:Softmax}
y_{i}=\frac{\exp(x_i)}{\sum\limits_{j}\exp(x_j)}.
\end{equation}
The training of the neural network can be performed in a supervised manner by labeling the training data. This defines a pairing of an input vector $\mathbf{s}_{0}$ and a desired output vector $\tilde{\mathbf{s}}_K$. Therefore, the training objective is to minimize, over the set of training inputs $\mathcal{S}$, the loss $\mathcal{L}(\boldsymbol{\theta})$, with respect to the parameter sets $\boldsymbol{\theta}$ of all $K$ layers, given by
\begin{equation}
\label{eq:Optimization}
\mathcal{L}(\boldsymbol{\theta})=\frac{1}{|\mathcal{S}|}\sum\limits_{(\mathbf{s}_{0,i},\tilde{\mathbf{s}}_{K,i})\in{\mathcal{S}}}\ell(f_{\textrm{ANN}}(\mathbf{s}_{0,i}),\tilde{\mathbf{s}}_{K,i})
\end{equation}
between an ANN output $\mathbf{s}_{K,i}=f_{\textrm{ANN}}(\mathbf{s}_{0,i})$ corresponding to the input $\mathbf{s}_{0,i}$ processed by all $K$ layers of the ANN,  and the desired, known output  $\tilde{\mathbf{s}}_{K,i}$. In \eqref{eq:Optimization}, $\ell(\mathbf{x},\mathbf{y})$ denotes the loss function and $|\mathcal{S}|$ denotes the cardinality of the training set (i.e., the size of the training set) containing 2-tuples $(\mathbf{s}_{0,i},\tilde{\mathbf{s}}_{K,i})$ of inputs and corresponding outputs. The loss function we consider in this work is the cross-entropy, defined as
\begin{equation}
\label{eq:CrossEntropy}
\ell(\mathbf{x},\mathbf{y})=-\sum\limits_{i}x_i\log(y_i).
\end{equation}
A common approach for optimization of the parameter sets $\boldsymbol{\theta}$ in \eqref{eq:Optimization}, which reduces computational demands, is to operate on a small \emph{batch} $\underline{\mathcal{S}}$ (called mini-batch) of the set of training data and perform the stochastic gradient descent (SGD) algorithm initialized with random $\boldsymbol{\theta}${\cite{Goodfellow}}, which is iteratively updated as 
\begin{equation}
\label{eq:SGD}
\boldsymbol{\theta}_{t}=\boldsymbol{\theta}_{t-1}-\eta\nabla\underline{\mathcal{L}}(\boldsymbol{\theta}_{t-1}),
\end{equation}
where $\eta$ is the learning rate of the algorithm and $\nabla\underline{\mathcal{L}}(\boldsymbol{\theta})$ is the gradient of the loss function of the mini-batch defined by
\begin{equation}
\label{eq:BatchLoss}
\underline{\mathcal{L}}(\boldsymbol{\theta})=\frac{1}{|\underline{\mathcal{S}}|}\sum\limits_{(\mathbf{s}_{0,i},\tilde{\mathbf{s}}_{K,i})\in{\underline{\mathcal{S}}}}\ell(f_{\textrm{ANN}}(\mathbf{s}_{0,i}),\tilde{\mathbf{s}}_{K,i}).
\end{equation} 
In modern deep learning, an efficient computation of the gradient in \eqref{eq:SGD} is achieved by error backpropagation~\cite{Goodfellow,Rumelhart}. A state-of-the-art algorithm with enhanced convergence is the Adam optimizer which dynamically adapts the learning rate~$\eta$~\cite{Kingma}. The Adam algorithm is used for optimization during the training process in this work. All numerical results in the manuscript have been generated using the deep learning library TensorFlow~\cite{Tensorflow}.

\section{Proposed End-to-End Communication System}\label{sec:system}
We implement the complete fiber-optic communication system and transmission chain including transmitter, receiver and channel as a complete end-to-end ANN, as suggested in~\cite{O'Shea,Doerner}. To show the concept, we focus on an IM/DD system, but we emphasize that the general method is not restricted to this scheme and can be easily extended to other, eventually more complex models. In the following we explain all the components of the transceiver chain as well as the channel model in detail. The full, end-to-end neural network chain is depicted in Fig.~\ref{fig:Schematic}.

\subsection{Transmitter Section}
We use a block-based transmitter as it has multiple advantages. Firstly, it is computationally simple, making it attractive for low-cost, high-speed implementations. Secondly, it allows massive parallel processing of the single blocks. Each block encodes an independent message $m\in\{1,\ldots, M\}$ from a set of $M$ total messages into a vector of $n$ transmit samples, forming a symbol. Each message represents an equivalent of $\log_2(M)$ bits. 

The encoding is done in the following way: The message $m$ is encoded into a \emph{one-hot vector} of size $M$, denoted as $\mathbf{1}_{m}\in{\mathbb{R}}^{M}$, where the $m$-th element equals 1 and the other elements are 0. Such one-hot encoding is the standard way of representing categorical values in most machine learning algorithms~\cite{Goodfellow} and facilitates the minimization of the symbol error rate. An integer encoding would for instance impose an undesired ordering of the messages. The one-hot vector is fed to the first hidden layer of the network, whose weight matrix and bias vector are $\mathbf{W}_{1}\in{\mathbb{R}}^{M\times 2M}$ and $\mathbf{b}_{1}\in{\mathbb{R}}^{2M}$, respectively. The second hidden layer has parameters $\mathbf{W}_{2}\in{\mathbb{R}}^{2M\times 2M}$ and $\mathbf{b}_{2}\in{\mathbb{R}}^{2M}$. The ReLU activation function~\eqref{eq:ReLuActivation} is applied in both hidden layers. The following layer prepares the data for transmission and its parameters are $\mathbf{W}_{3}\in{\mathbb{R}}^{2M\times n}$ and $\mathbf{b}_{3}\in{\mathbb{R}}^{n}$, where $n$ denotes the number of waveform samples representing the message. The dimensionality of this layer determines the oversampling rate of the transmitted signal. In our work, $4\times$ oversampling is considered and thus the message is effectively mapped onto a symbol of $n/4$ samples. As fiber dispersion introduces memory between several consecutive symbols, multiple transmitted blocks need to be considered to model realistic transmission. Hence, the output samples of $N$ neighboring blocks (that encode potentially different inputs) are concatenated by the serializer to form a sequence of $N\cdot n$ samples ready for transmission over the channel. All these $N$ ANN blocks have identical weight matrices and bias vectors. The system can be viewed as an \emph{autoencoder} with an effective information rate $R=\log_{2}(M)$ bits/symbol. We consider unipolar signaling and the ANN transmitter has to limit its output values to the Mach-Zehnder modulator (MZM) relatively linear operation region $[0;\pi/4]$. This is achieved by applying the clippling activation function for the final layer which combines two ReLUs as follows
\begin{equation}
\label{eq:ModifiedActivation}
\alpha_{\text{Clipping}}(\mathbf{x})=\alpha_{\tiny{\text{ReLU}}}\left(\mathbf{x}-\epsilon\right)-\alpha_{\tiny{\text{ReLU}}}\left(\mathbf{x}-\frac{\pi}{4}+\epsilon\right),
\end{equation}
where the term $\epsilon=\sigma_{\text{q}}/2$ ensures the signal is within the MZM limits after quantization noise is added by the DAC. The variance $\sigma^2_{\text{q}}$ of the quantization noise is defined below.   

\subsection{Communication Channel}
The main limiting factor in IM/DD systems is the intersymbol interference (ISI) as a result of optical fiber dispersion~\cite{Agrawal}. Moreover, in such systems, simple photodiodes (PDs) are used to detect the intensity of the received optical field and perform opto-electrical conversion, so called square-law detection. As a consequence of the joint effects of dispersion and square-law detection, the IM/DD communication channel is nonlinear and has memory. 

In our work, the communication channel model includes low-pass filtering (LPF) to account for the finite bandwidth of transmitter and receiver hardware, DAC, ADC, MZM, photo-conversion by the PD, noise due to amplification and optical fiber transmission. The channel is considered part of the system implemented as an end-to-end deep feedforward neural network shown in Fig.~\ref{fig:Schematic}. The signal that enters the section of the ANN after channel propagation can be expressed as (neglecting the receiver LPF for ease of exposition)
\begin{equation}
\label{eq:Channel}
r(t)= |u(t)|^{2} + n_{\text{Rec.}}(t),
\end{equation}
where $u(t)=\hat{h}\{x(t)\}$ is the waveform after fiber propagation, $x(t)$ is the transmit signal, $\hat{h}\{\cdot\}$ is an operator describing the effects of the electrical field transfer function of the modulator and the fiber dispersion, $n_{\text{Rec.}}(t)$ is additive Gaussian noise arising, e.g., from the trans-impedance amplifier (TIA) circuit. We select the variance of the noise to match the signal-to-noise ratios (SNRs) after photodetection obtained in our experimental setup. Further details on the  SNR values at the examined distances are presented below in Sec.~\ref{sec:experiment}. We now discuss in more detail the system components.

Chromatic dispersion in the optical fiber is mathematically expressed by the partial differential equation \cite{Agrawal}
\begin{equation}
\label{eq:Dispersion}
\frac{\partial A}{\partial z}=-j\frac{\beta_{2}}{2}\frac{\partial^{2}A}{\partial t^{2}}\,,
\end{equation}
where $A$ is the complex amplitude of the optical field envelope, $t$ denotes time, $z$ is the position along the fiber and $\beta_{2}$ is the dispersion coefficient. Equation \eqref{eq:Dispersion} can be solved analytically in the frequency domain by taking the Fourier transform, yielding the dispersion frequency domain transfer function
\begin{equation}
\label{eq:DispersionFreq}
\mathcal{D}(z,\omega)=\exp\left(j\frac{\beta_{2}}{2}\omega^2z\right),
\end{equation}
where $\omega$ is the angular frequency. In our work, fiber dispersion is applied in the frequency domain on the five-fold zero-padded version of the signal stemming from $N$ concatenated blocks. The FFT and IFFT necessary for conversion between time and frequency domain form part of the ANN and are provided by the TensorFlow library~\cite{Tensorflow}.

The MZM is modeled by its electrical field transfer function, a sine which takes inputs in the interval $[-\pi/2;\pi/2]$ \cite{Napoli}. This is realized in the ANN by using a layer that consists just of the MZM function $\alpha_{\text{MZM}}(\mathbf{x}) =\sin(\mathbf{x})$, where the sine is applied element-wise. The DAC and ADC components introduce additional quantization noise due to their limited resolution. We model this noise $n_{\text{DAC}}(t)$ and $n_{\text{ADC}}(t)$ as additive, uniformly distributed noise with variance determined by the effective number of bits (ENOB) of the device~\cite{Pearson}
\begin{equation}
\label{eq:ENOB}
\sigma_{q}^2=3P\cdot 10^{-(6.02\cdot\textnormal{ENOB}+1.76)/10},
\end{equation}
where $P$ is the average power of the input signal. Low-pass filtering is applied before the DAC/ADC components to restrict the bandwidth of the signal. Note that both LPF stages and the chromatic dispersion stage can be modeled as purely linear stages of the ANN, i.e., a multiplication with a correspondingly chosen matrix $\mathbf{W}_k$. The MZM and PD stages are modeled by a purely nonlinear function $\alpha_k$.

\subsection{Receiver Section}
After square-law detection, amplification, LPF, and ADC, the central block is extracted for processing in the receiver section of the neural network. The architecture of the following layers is identical to those at the transmitter side in a reverse order. The parameters of the first receiver layer are $\mathbf{W}_{4}\in{\mathbb{R}}^{n\times 2M}$, $\mathbf{b}_{4}\in{\mathbb{R}}^{2M}$ with ReLU activation function~\eqref{eq:ReLuActivation}. The next layer has parameters $\mathbf{W}_{5}\in{\mathbb{R}}^{2M\times 2M}$, $\mathbf{b}_{5}\in{\mathbb{R}}^{2M}$, also with ReLU activation function. The parameters of the final layer in the ANN are $\mathbf{W}_{6}\in{\mathbb{R}}^{2M\times M}$ and $\mathbf{b}_{6}\in{\mathbb{R}}^{M}$. The final layer's activation is the \textit{softmax} function~\eqref{eq:Softmax} and thus the output is a probability vector $\mathbf{y}\in{\mathbb{R}}^{M}$ with the same dimension as the one-hot vector encoding of the message. At this stage, a decision on the transmitted message is made and a block (symbol) error occurs when $m\neq \argmax(\mathbf{y})$, where $m$ is the index of the element equal to 1 in the one-hot vector ($\mathbf{1}_{m}$) representation of the input message. Then the block error rate (BLER) can be estimated as
\begin{equation}
\label{eq:BLER}
\text{BLER}=\frac{1}{|\mathcal{S}|}\sum\limits_{i\in{\mathcal{S}}}\mathop{\mathbbm{1}}\left\{m_{i}\neq\argmax(\mathbf{y}_{i})\right\},
\end{equation}
where $|\mathcal{S}|$ is the cardinality of the set of messages $\mathcal{S}$ and $\mathbbm{1}$ is the indicator function, equal to 1 when the condition in the brackets is satisfied and 0 otherwise.

In our work, the bit-error rate (BER) is examined as an indicator of the system performance. For computing the BER, we use an ad hoc bit mapping by assigning the Gray code to the input $m\in\{1,\ldots, M\}$. Whenever a block is received in error, the number of wrong bits that have occurred are counted. Note that this approach is sub-optimal as the deep learning algorithm will only minimize the BLER and a symbol error may not necessarily lead to a single bit error. In our simulation results, we will hence provide a lower bound on the achievable BER with an optimized bit mapping by assuming that at most a single bit error occurs during a symbol error.

Note that the structure we propose is only able to compensate for chromatic dispersion within a block of $n$ receiver samples, as there is no connection between neighboring blocks. The effect of dispersion from neighboring blocks is treated as extra noise. The block size $n$ (and $m$) will hence limit the achievable distance with the proposed system. However, we could in principle extend the size of the receiver portion of the ANN to jointly process multiple blocks to dampen the influence of dispersion. This will improve the resilience to chromatic dispersion at the expense of higher computation complexity.

\subsection{Training}
The goal of the training is to obtain an efficient \emph{autoencoder}~\cite[Ch.~14]{Goodfellow}, i.e., the output of the final ANN softmax layer should be ideally identical to the one-hot input vector. Such an autoencoder will minimize the end-to-end BLER. In this work, the ANN is trained with the Adam optimizer~\cite{Kingma} on a set of $|\mathcal{S}|\!=\!25\!\cdot\!10^{6}$ randomly chosen messages (and messages of the neighboring transmit blocks) and mini-batch size $|\underline{\mathcal{S}}|\!=\!250$, corresponding to 100 000 iterations of the optimization algorithm. {It is worth noting that in most cases, convergence in the loss and validation symbol error rate of the trained models was obtained after significantly less than 100 000 iterations, which we used as a fixed stopping criterion.} During training, noise is injected into the channel layers of the ANN, as shown in Fig.~\ref{fig:Schematic}. A truncated normal distribution with standard deviation $\sigma=0.1$ is used for initialization of the weight matrices $\mathbf{W}$. The bias vectors $\mathbf{b}$ are initialized with 0. Validation of the training is performed during the optimization process every 5000 iterations. The validation set has the size $|\mathcal{S}_v|\!=\!15\cdot10^6$.  Good convergence of the validation BLER and the corresponding BER is achieved. The trained model is saved and then loaded separately for testing which is performed over a set of different $|\mathcal{S}_t|\!=\!15\cdot10^8$ random input messages. The BER results from testing are shown in the figures throughout this manuscript. We have confirmed the convergence of the results as well for mini-batch sizes of $|\underline{\mathcal{S}}|\!=\!125$ and $500$, and also when the training set was increased to $|\mathcal{S}|\!=\!50\!\cdot\!10^{6}$.

{When designing ANNs, the choice of hyper-parameters such as the number of layers, number of nodes in a hidden layer, activation functions, mini-batch size, learning rate, etc.  is important. The optimization of the hyper-parameters was beyond the scope of our investigation. In this work they were chosen with the goal to keep the networks relatively small and hence the training effort manageable. Better results in terms of performance and its trade-off with complexity can be obtained with well-designed sets of hyper-parameters.}
\section{System Performance}\label{sec:simulation}
\begin{figure}[tb!]
\centering
\includegraphics[width=\columnwidth]{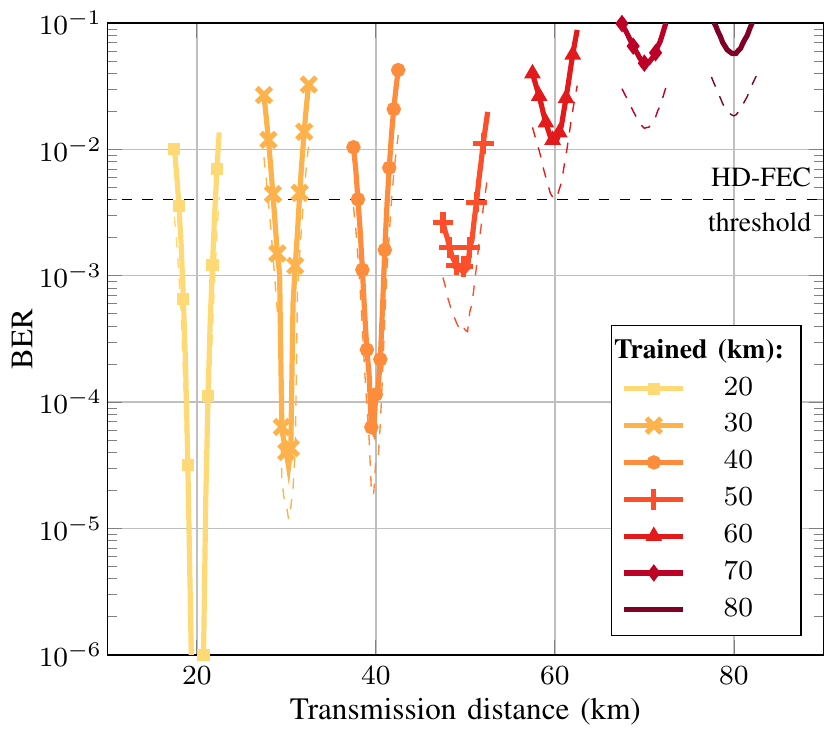}
\caption{Bit error rate as a function of transmission distance for systems trained at a fixed nominal distance of ($20+i\cdot 10$)\,km, with $i\in\{0,\ldots,6\}$. The horizontal dashed line indicates the 6.7\% HD-FEC threshold. Thin dashed lines below the curves give a lower bound on the achievable BER when optimal bit mapping, such that a block error results in a single bit error, is assumed.}
\label{fig:at_fixed_nominal}
\end{figure}

Table \ref{table1:Network parameters} lists the simulation parameters for the end-to-end deep-learning-based optical fiber system under investigation. We assume a set of $M=64$ input messages which are encoded by the neural network at the transmitter into a symbol of 48 samples at 336 GSa/s in the simulation. {This rate corresponds to the 84\,GSa/s sampling rate of the DAC used in experiment multiplied by the oversampling factor of 4, which we assume in simulation.} The bandwidth of the signal is restricted by {a 32\,GHz low-pass filter} to account for the significantly lower bandwidth of today's hardware. Thus the information rate of the system becomes $R=6$\,bits/sym. Symbols are effectively transmitted at 7\,GSym/s and thus the system operates at a bit rate of 42\,Gb/s.
 \begin{table}
 \caption{Simulations Parameters}
 \label{table1:Network parameters}
 \centering
 \begin{tabular}{cc}
 \toprule
 Parameter & Value \\
 \midrule
 \textit{M} & 64   \\
 \textit{n} & 48 \\
 \textit{N} & 11  \\
 Oversampling & 4 \\
 Sampling rate & 336 GSa/s \\
 Symbol rate & 7 GSym/s \\
 Information rate & 6 bits/symbol \\
 LPF bandwidth & 32 GHz \\
 DAC/ADC ENOB & 6\\
 Fiber dispersion parameter & 17 ps/nm/km \\
 \bottomrule
 \end{tabular}
\end{table}
Figure~\ref{fig:at_fixed_nominal} shows the BER performance at different transmission distances. For this set of results, the ANN was trained for 7 different distances in the range 20 to 80\,km in steps of $10$\,km and the distance was kept constant during training. During the testing phase, the distance was swept. BERs below the 6.7\% hard decision FEC (HD-FEC) threshold of $4\cdot10^{-3}$ are achieved at all examined distances between 20 and 50\,km. Moreover, up to 40\,km the BER is below $10^{-4}$. Systems trained at distances longer than 50\,km achieve BERs above $10^{-2}$. The figure also displays the lower bound on the achievable BER for each distance. This lower bound is obtained by assuming that a block error gives rise to a single bit error. An important observation is that the lowest BERs are obtained at the distances for which the system was trained and there is a rapid increase in the BER when the distance changes. Such a behavior is a direct consequence of the implemented training approach which optimizes the system at a particular distance without any incentive of robustness to variations. As the amount of dispersion changes with distance, the optimal neural network parameters differ accordingly and thus the BER increases as the distance changes. We therefore require a different optimization method that yields ANNs that are robust to distance variations and hence offer new levels of flexibility.

\begin{figure}[tb!]
\centering
\includegraphics[width=\columnwidth]{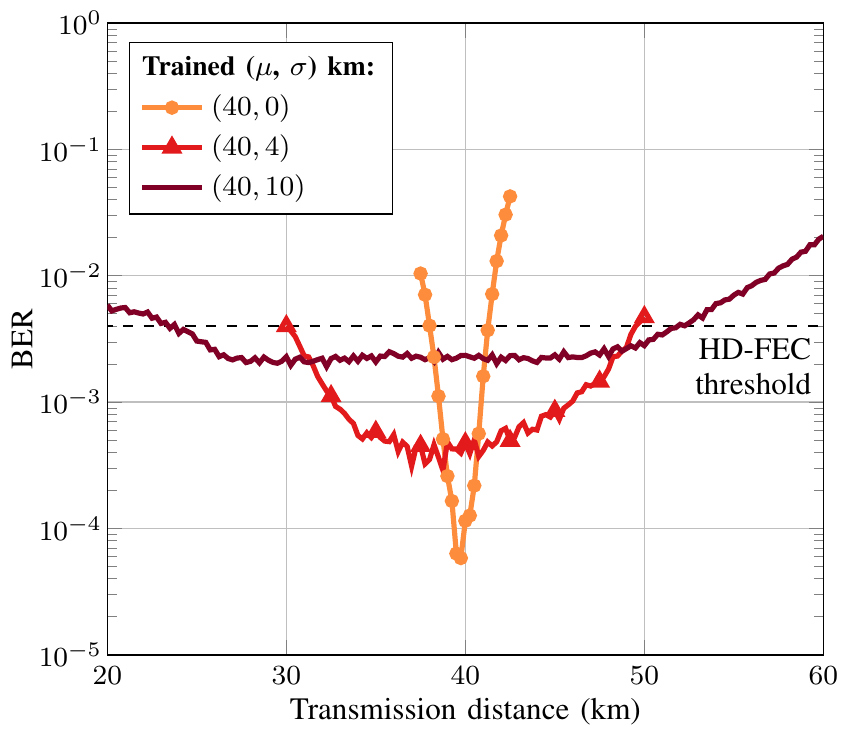}
\caption{ Bit error rate as a function of transmission distance for systems where the training is performed at normally distributed distances with mean $\mu$ and standard deviation $\sigma$. The horizontal dashed line indicates the 6.7\% HD-FEC threshold.}
\label{fig:at_normal_distributed_40km}
\end{figure}

To address these limitations of the training process, we train the ANN in a process where instead of fixing the distance, the distance for every training message is randomly drawn from a Gaussian distribution with a mean~$\mu$ and a standard deviation~$\sigma$. During optimization, this allows the deep learning to converge to more generalized ANN parameters, robust to certain variation of the dispersion. Figure~\ref{fig:at_normal_distributed_40km} shows the test BER performance of the system trained at a mean distance $\mu=40$\,km and different values of the standard deviation. We see that for both cases of $\sigma=4$ and $\sigma=10$ this training method allows BER values below the HD-FEC threshold in wider ranges of transmission distances than for $\sigma=0$. For instance, when $\sigma=4$, BERs below the $4\cdot10^{-3}$ threshold are achievable between 30.25\,km and 49.5\,km, yielding a range of operation of 19.25\,km. The distance tolerance is further increased when $\sigma=10$ is used for training. In this case, the obtained BERs are higher due to the compromise taken, but still below the HD-FEC threshold for a range of 27.75\,km, between 24\,km up to 51.75\,km. A practical implementation of the proposed fiber-optic system design is expected to greatly benefit from such a training approach as it introduces both robustness and flexibility of the system to variations in the link distance. As a consequence of generalizing the learning over varied distance, the minimum achievable BERs are higher compared to the system optimized at a fixed distance, presented in Fig.~\ref{fig:at_fixed_nominal}, and there exists a trade-off between robustness and performance.

\begin{figure}[tb!]
\centering
\includegraphics[width=\columnwidth]{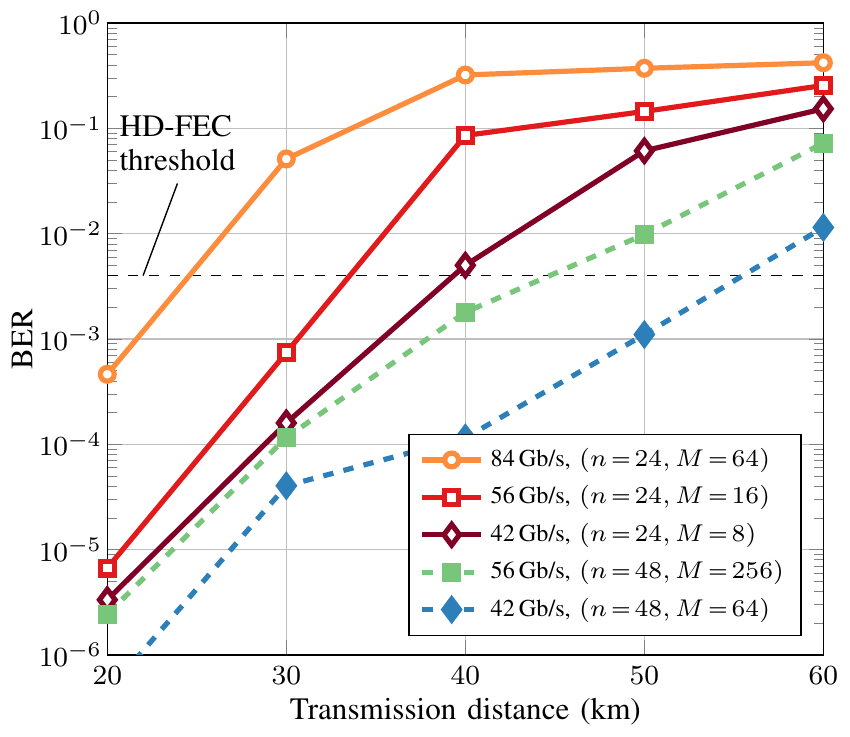}
\caption{Bit error rate as a function of transmission distance for systems with different information rates. The training is performed at a fixed nominal distance. }
\label{fig:Rates}
\end{figure}

\begin{figure*}[ht]
\centering
     \includegraphics[width=\textwidth]{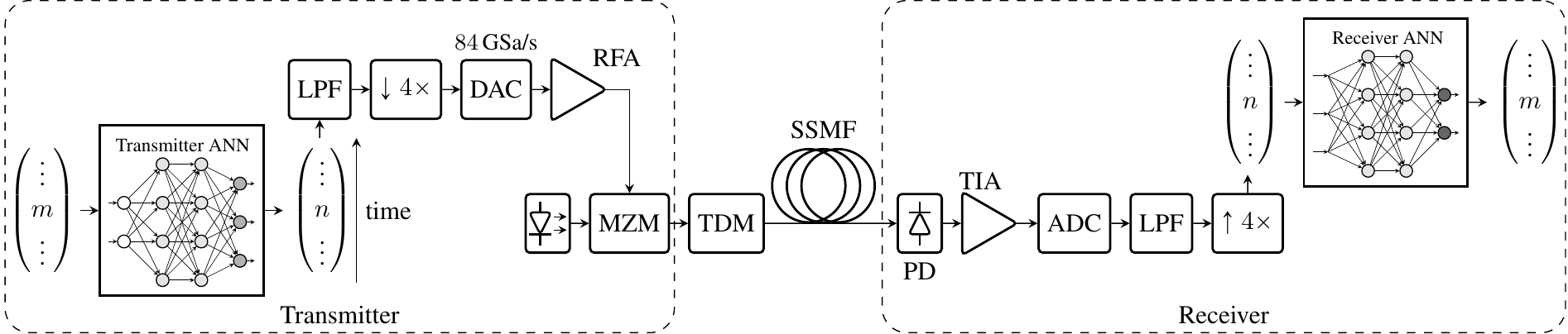}
      \caption{Schematic of the experimental setup for system validation.}
       \label{fig:ExperimentSchematic}
\end{figure*}

So far we examined an end-to-end deep learning optical fiber system where an input message carrying 6 bits of information ($M=64$) is encoded into a band-limited symbol of 48 samples ($n=48$ with an oversampling factor of 4) at 336\,GSa/s. Thus, the result is an autoencoder operating at the bit rate of 42\,Gb/s. In the following, we examine different rates by varying the size of $M$ and $n$ and thus the size of the complete end-to-end neural network. For this investigation, we fixed the sampling rate of the simulation to 336\,GSa/s. In Figure~\ref{fig:Rates} solid lines show the BER performance of the system at different rates when the number of symbols used to encode the input message is decreased, in particular we use $n=24$, thus yielding a symbol rate of 14\,GSym/s. In such a way bit rates of 42\,Gb/s, 56\,Gb/s and 84\,Gb/s are achieved for $M=8$, $M=16$, and $M=64$, respectively. We see that the BER at 84\,Gb/s rapidly increases with distance and error rates below the HD-FEC can be achieved only up to 20\,km. On the other hand, 42\,Gb/s and 56\,Gb/s can be transmitted reliably at 30\,km. An alternative to decreasing the transmitted samples in a block is to increase the information rate of the system by considering input messages with a larger information content. Dashed lines in Fig.~\ref{fig:Rates} show the cases of $M=64$, $n=48$ and $M=256$, $n=48$, corresponding to bit rates of 42\,Gb/s and 56\,Gb/s. In comparison to the case where $n=24$, such systems have an extended operational reach below the BER threshold, due to the larger block size and the reduce influence of chromatic dispersion. For example, the 56\,Gb/s system can achieve BER below the HD-FEC at 40\,km, while for 42\,Gb/s, this distance is 50\,km. Thus increasing the information rates by assuming larger $M$ enables additional reach of 10\,km and 20\,km at 56\,Gb/s and 42\,Gb/s, respectively. However, a drawback of such a solution is the larger ANN size, thus increasing the computational and memory demands as well as training times. Figure~\ref{fig:Rates} shows that the general approach of viewing the optical fiber communication system as a complete end-to-end neural network can be applied for designing systems with different information rates and gives an insight on the possible implementation approaches.

\section{Experimental Validation}\label{sec:experiment}

To complement the simulation results, we built an optical transmission system to demonstrate and validate experimentally the results obtained for the end-to-end deep learning IM/DD system operating at 42\,Gb/s. Moreover, we utilize the proposed training method and train our models at the examined distances of 20, 40, 60, or 80\,km with a standard deviation of $\sigma=4$. Figure~\ref{fig:ExperimentSchematic} illustrates the experimental setup. The SNRs after photodetection assumed in the end-to-end training process during generation of the transmit waveforms are 19.41\,dB, 6.83\,dB, 5.6\,dB and 3.73\,dB at 20, 40, 60 and 80\,km, respectively, corresponding to measured values for the 42\,Gbaud PAM2 system, which is described in this section and used for comparison reasons. Since the training for the experiment is performed at distances with a certain standard deviation, linear interpolation is used to find the SNR values at distances different from the above.

The transmit waveforms were obtained by feeding a random sequence to the transmitter ANN, filtering by a LPF with 32\,GHz bandwidth, downsampling and DAC (after standard {linear finite-impulse response (FIR)} DAC pre-emphasis). In the experiment, we downsample by a factor of 4 the resulting filtered concatenated series of symbols, each now containing 12 samples. Because of LPF, there is no loss of information, since the original series of symbols, at 48 samples each and running at 336 GSa/s, can be exactly regenerated from this downsampled series of symbols, 12 samples per symbol at 84 GSa/s. The waveform is then used to modulate an MZM, where the bias point is meticulously adjusted to match the one assumed in simulations. The optical signal at 1550\,nm wavelength is propagated over a fixed fiber length of 20, 40, 60, or 80 km and through a Tunable Dispersion Module (TDM), which is deployed to allow sweeping the dispersion around a given value. The received optical waveform is direct detected by a PIN+TIA and real-time sampled and stored for the subsequent digital signal processing. There is no optical amplification in the testbed. After synchronization, proper scaling and offset of the digitized photocurrent, the upsampled received waveforms are fed block-by-block to the receiver ANN. After fine-tuning of the receiver ANN parameters, the BLER and BER of the system are evaluated. In the experiment, $40\cdot10^{6}$  blocks are transmitted and received for each dispersion value. This is achieved by transmitting 1000 sequences of $40\cdot10^{3}$ blocks. To compare our system with conventional IM/DD schemes operating at 42\,Gb/s, we perform experiments at the examined distances for two reference systems: the first operating at 42\,Gbaud with PAM2 and raised cosine pulses (roll-off of 0.99); the second operating at 21\,Gbaud with PAM4 and raised cosine pulses (roll-off of 0.4). Both reference system use  feedforward equalization (FFE) with 13 taps (T/2-spaced) at the receiver. {It is easy to see that the computational complexity of this simple linear equalization scheme is lower than the complexity of a deep ANN-based receiver. Nevertheless, we use the comparison to emphasize on the viability of implementing the optical fiber system as an end-to-end deep ANN. Hence, possible complexity reductions in the design are beyond the scope of the manuscript.} 

While carrying out the experiment, we found that the ANN trained in the simulation was not fully able to compensate distortions from the experimental setup. Hence, we decided to retrain the receiver ANN (while keeping the transmitter ANN fixed) to account for the experimental setup. Retraining has been carried out for every measured distance. For the retraining of the receiver ANN, we used a set of $|\mathcal{S}|\!=\!30\cdot10^{6}$ {(75\% of all block traces)} received blocks, while validation during this process is performed with a set of $|\mathcal{S}_v|\!=\!5\cdot10^6$ {(12.5\% of all block traces)} different blocks (from different measurements). The fine-tuned model is tested over the remaining  $|\mathcal{S}_t|\!=\!5\cdot10^6$ {(12.5\% of all block traces)} (these were \emph{not} used for training and validation). {The subdivision of the experimental data into training, validation and testing sets is in accordance to the guidelines given in ~\cite[Sec. 5.3]{Goodfellow}.} Training was carried out over 4 epochs over the experimental data, which was enough to see good convergence. {In a single epoch each of the received blocks for training is used once in the optimization procedure, yielding a single \emph{pass} of the training set $|\mathcal{S}|$ through the algorithm. Realization of 4 epochs improved convergence and further ensured that we perform enough training iterations to observe convergence (see Sec.~\ref{sec:system}-D).} For retraining the receiver ANN, the layer parameters are initialized with the values obtained in simulation prior to the experiment. The output of the receiver ANN is optimized with respect to the labeled experimental transmit messages, following the same procedure as described in Sec.~\ref{sec:ANN}. Again, a mini-batch size of $|\underline{\mathcal{S}}|\!=\!250$ has been used. Experimental BER results are then obtained on the testing set only and are presented in what follows.

\begin{figure}[tb!]
\centering
\includegraphics[width=\columnwidth]{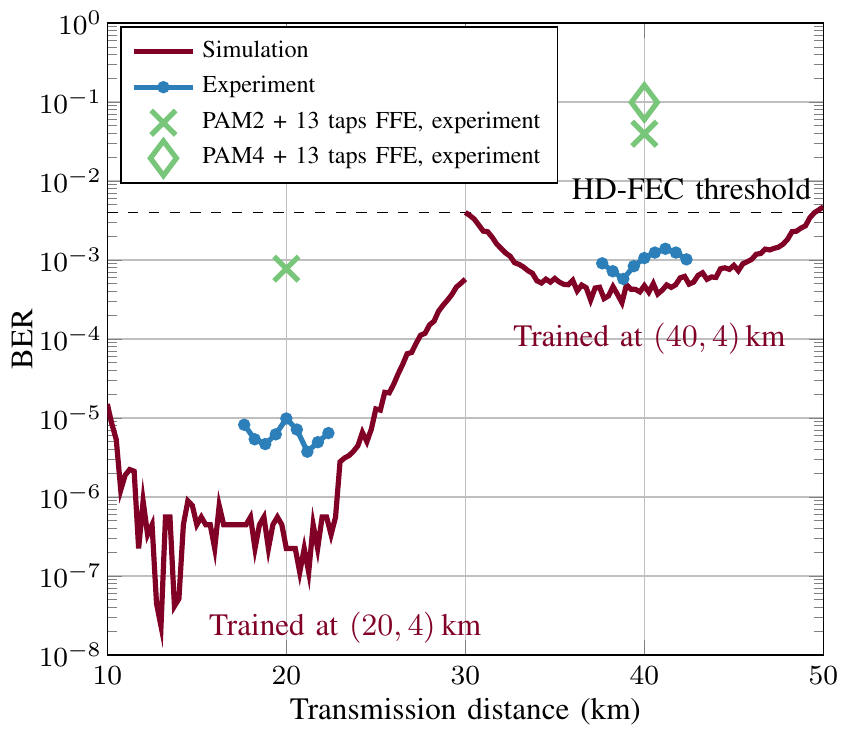}
\caption{Experimental BER performance for systems trained at $(20,4)$\,km and $(40,4)$\,km. }
\label{fig:Experiment_20_40}
\end{figure}

\begin{figure}[tb!]
\centering
\includegraphics[width=\columnwidth]{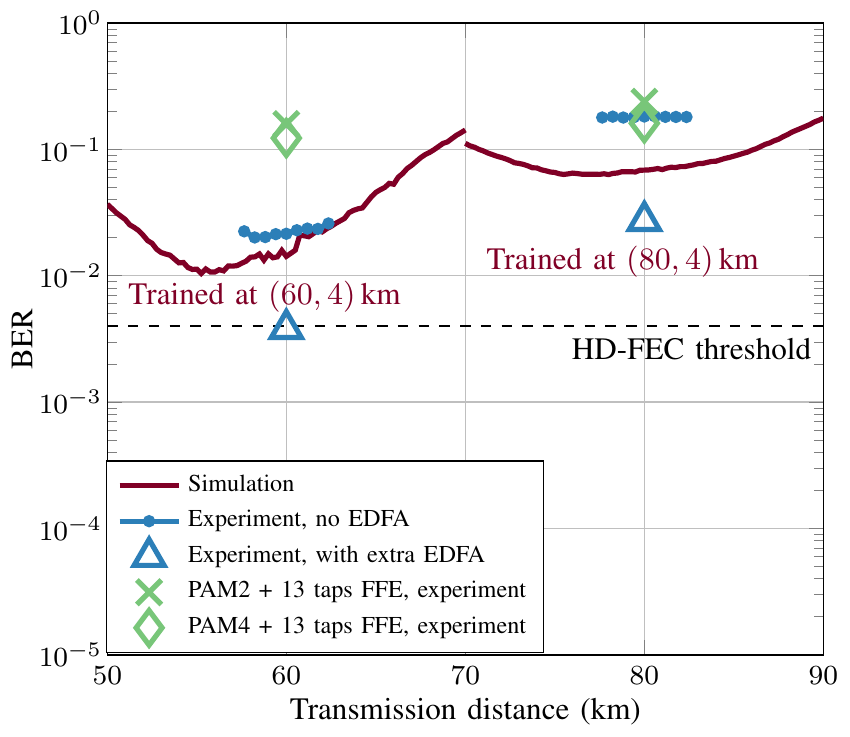}
\caption{Experimental BER performance for systems trained at $(60,4)$\,km and $(80,4)$\,km. }
\label{fig:Experiment_60_80}
\end{figure}

Figure~\ref{fig:Experiment_20_40} shows the experimental results for a fiber of length 20\,km and 40\,km. The TDM dispersion value was swept between $-40$\,ps and $+40$\,ps, resulting in effective link distances in the ranges of $17.65-22.35$\,km and $37.65-42.35$\,km, respectively. For the system around 20\,km, BERs below $10^{-5}$ have been  achieved experimentally at all distances. In particular, the lowest BER of $3.73\cdot10^{-6}$ has been obtained at 21.18\,km. For comparison, the PAM2 system experimentally achieves $7.77\cdot10^{-4}$ BER at 20\,km and is therefore significantly outperformed by the end-to-end deep learning optical system. At 40\,km, the proposed system outperforms both the 42\,Gbaud PAM2 and the 21\,Gbaud PAM4 schemes, as neither of these can achieve BERs below the HD-FEC threshold. On the other hand, the ANN-based system achieved BERs below $1.4\cdot10^{-3}$ at all distances in the examined range. In particular, BERs of $1.05\cdot10^{-3}$ at 40\,km and a lowest BER of $5.75\cdot10^{-4}$ at 38.82\,km have been obtained. Furthermore, we see that both sets of experimental results at 20\,km and at 40\,km are in excellent agreement with the simulation results.

Figure~\ref{fig:Experiment_60_80} shows the experimental results at 60\,km and 80\,km fiber length and TDM dispersion swiped between $-40$\,ps and $+40$\,ps, yielding effective link distances in the ranges $57.65-62.35$\,km and $77.65-82.35$\,km, respectively. For both systems we see that BERs below the HD-FEC threshold cannot be achieved by the end-to-end deep learning approach, as predicted by the simulation. Nevertheless, at 60\,km the system still outperforms the PAM2 and PAM4 links. However, for the 80\,km, link the thermal noise at the receiver becomes more dominant due to the low signal power levels without optical amplification. In combination with the accumulated dispersion, whose effects at 80\,km extend across multiple blocks and cannot be compensated by the block-by-block processing, this results in operation close to the sensitivity limits of the receiver which ultimately restricts the achievable BERs. 
\begin{figure*}[ht]
\centering
\begin{tabular}{c}
\includegraphics{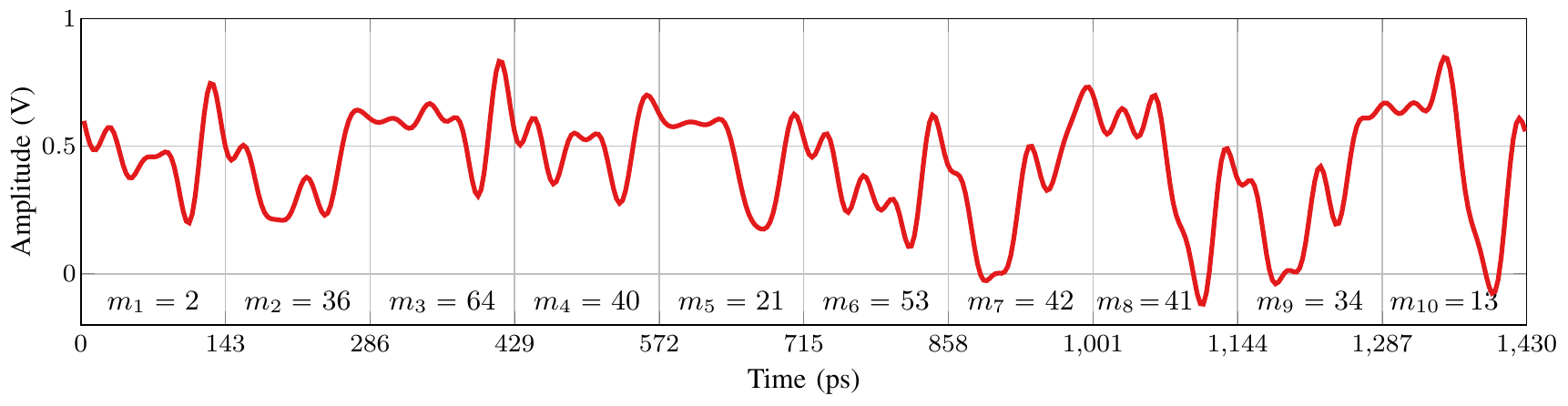}\\
\includegraphics{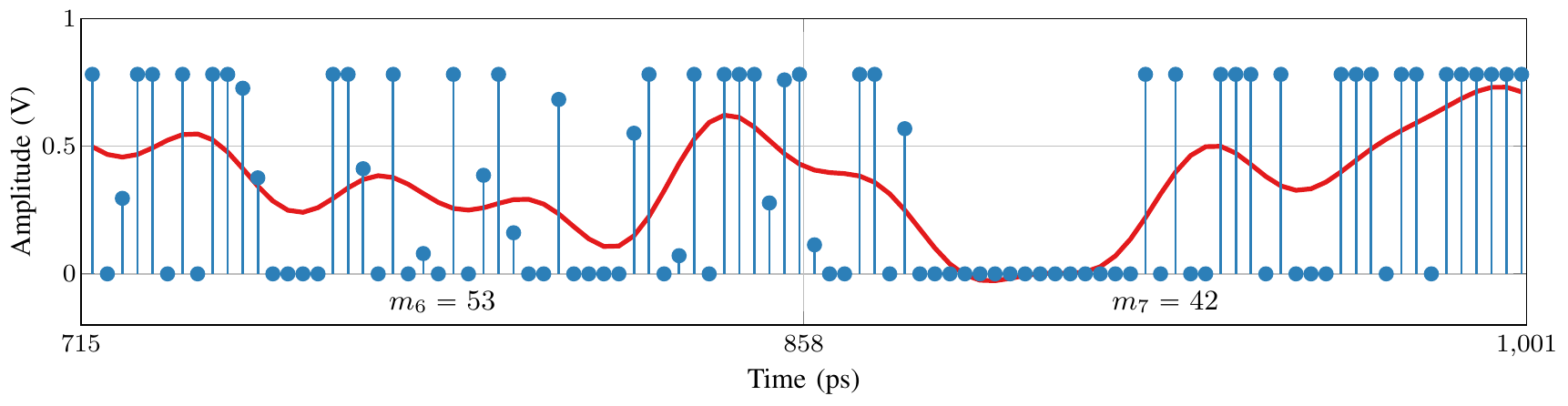}
\end{tabular}
\caption{{Top: Output of the transmitter ANN, trained at (40,4)\,km, after filtering with 32\,GHz Brickwall LPF for the representative random sequence of 10 symbols $(m_t)_{t=1}^{10} = (2,36,64,40,21,53,42,41,34,13)$ transmitted at 7\,GSym/s, i.e. $T\approx143$\,ps. Bottom: Un-filtered ANN output samples, 48 per symbol, for the sub-sequence $(m_t)_{t=6}^{7} =(53,42)$.}}
\label{fig:Waveform_LP}
\end{figure*}

To further investigate the impact of received signal power on the performance of the system, we included an erbium-doped fiber amplifier (EDFA) in the deep learning-based test-bed for pre-amplification at the receiver. Thereby, the received power is increased from -13 and -17\,dBm at 60\,km and 80\,km, respectively to -7\,dBm. The obtained BERs at these distances are shown as well in Fig.~\ref{fig:Experiment_60_80}. We see that by changing the link to include an extra EDFA, the end-to-end deep learning system achieves significantly improved performance. In particular, at 60\,km, a BER of $3.8\cdot10^{-3}$, slightly below the HD-FEC threshold, can be achieved. Due to dispersion and block-based processing, there is a significant impact at 80\,km as well, where the obtained BER is $2.8\cdot10^{-2}$. These results highlight the great potential for performance improvement by including different link configurations inside the end-to-end learning process.

{\section{Discussion}\label{sec:discussion}}

\begin{figure*}
\includegraphics[width=\textwidth]{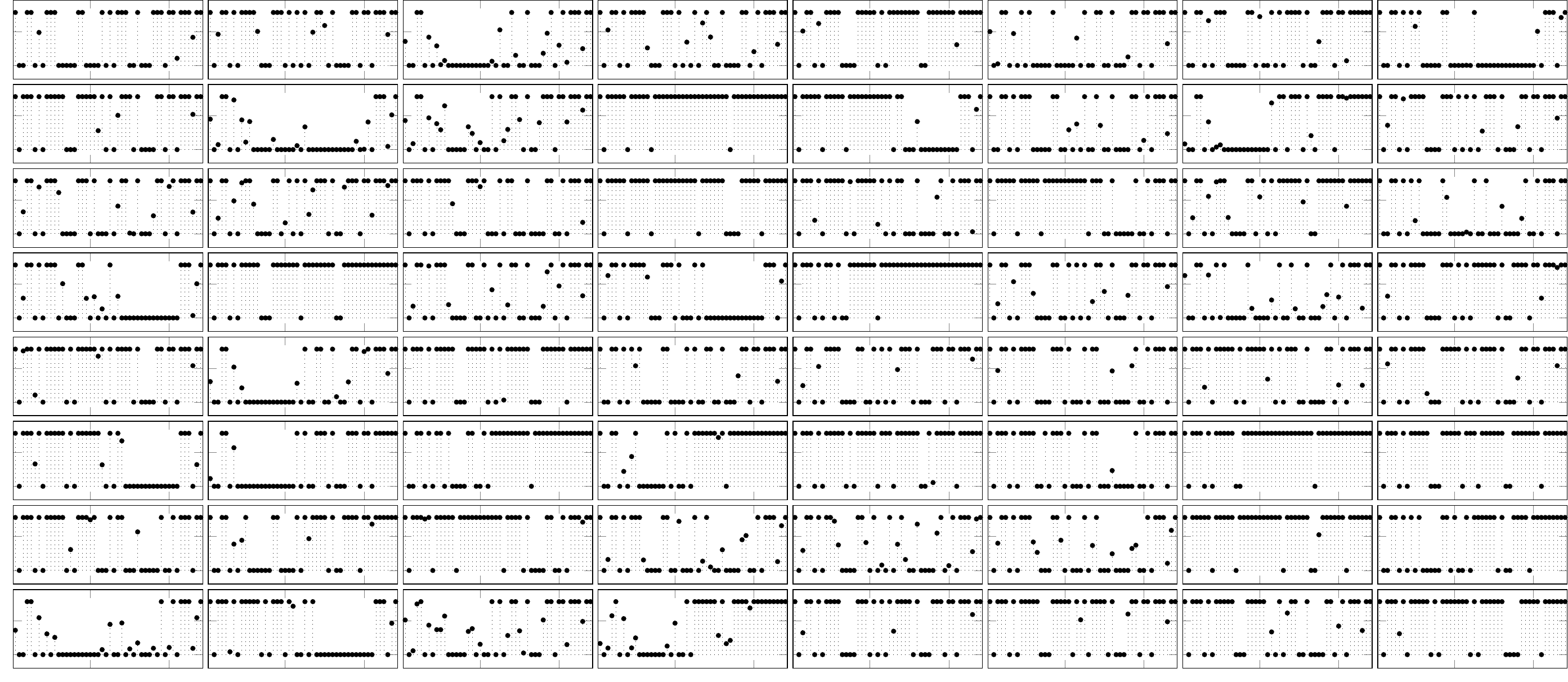}
\caption{{All 64 possible outputs ($m=1$ to $m=64$, upper left to bottom right) of the transmitter ANN before low-pass filtering.}}
\label{fig:Tx_mapping}
\end{figure*}

\begin{figure}[tb!]
\centering
\includegraphics[width=\columnwidth]{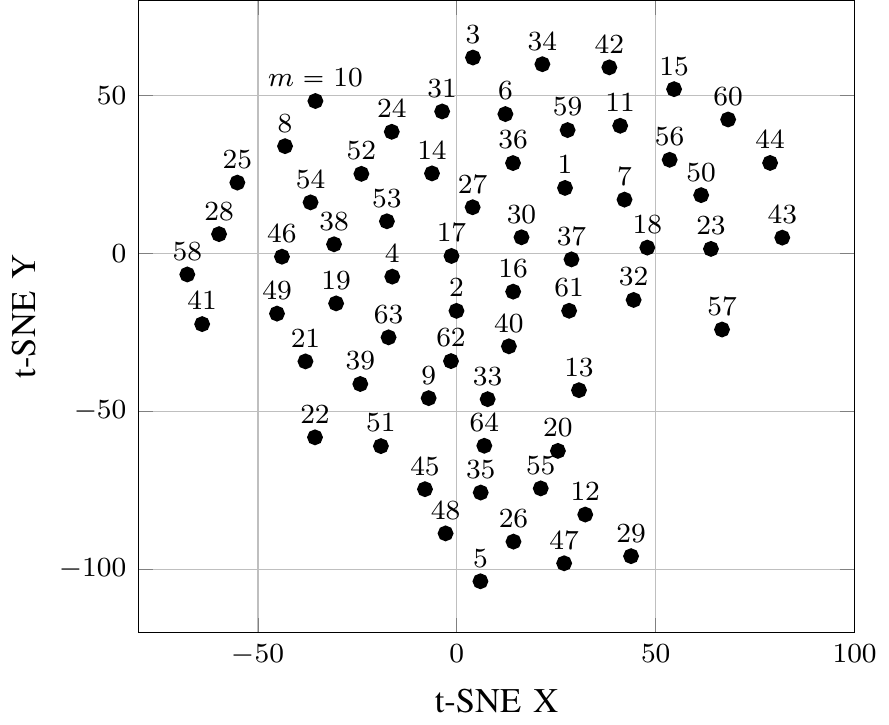}
\caption{{t-SNE representation of the multi-dimensional waveforms output of the transmitter ANN on the two-dimensional plane. The points are labeled with their respective message number $m$.}}
\label{fig:tSNE_constellation}
\end{figure}

\begin{figure}[tb!]
\centering
\includegraphics[width=\columnwidth]{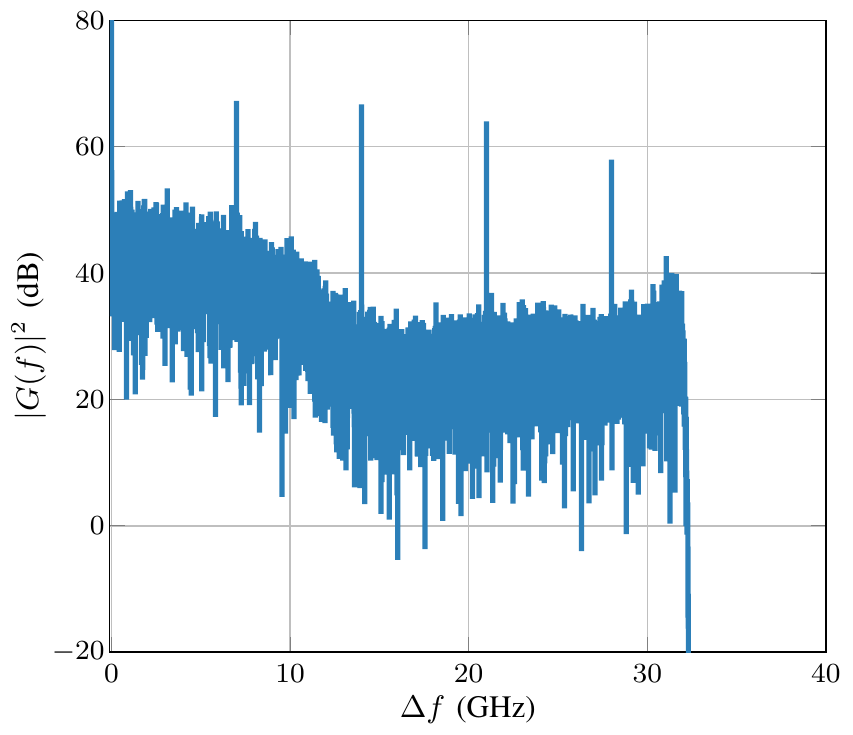}
\caption{{Spectrum of the 32\,GHz Brickwall low-pass filtered waveform at the output of the transmitter ANN, trained at (40,4)\,km.}}
\label{fig:Spectrum}
\end{figure}

{\subsection{Transmitted Signal Characteristics}}
{In our end-to-end optimization of the transceiver, the transmitter learns waveform representations which are robust to the optical channel impairments. In the experiment, we apply random sequences to the transmitter ANN, followed by 32\,GHz LPF to generate the transmit waveforms. We now exemplify the temporal and spectral representations of the transmit signal. Figure~\ref{fig:Waveform_LP} (top) shows the filtered output of the neural network, trained at (40,4)\,km, for the representative 10-symbol message  sequence $(m_t)_{t=1}^{10} = (2,36,64,40,21,53,42,41,34,13)$, with $m_t\in\{1,\ldots,64\}$ denoting the input message to the ANN at time/block $t$. Each symbol carries 6\,bits of information, consists of 48 samples, and is transmitted at 7\,GSym/s, yielding a symbol duration $T\approx 143$\,ps. We observe that, as an effect of the clipping layer in our transmitter ANN, the waveform amplitude is limited in the linear region of operation of the Mach-Zehnder modulator with small departure from the range $[0;\frac{\pi}{4}]$ due to the filtering effects. Figure~\ref{fig:Waveform_LP} (bottom) also shows the un-filtered 48 samples for each symbol in the sub-sequence $(m_t)_{t=6}^7=(53,42)$. These blocks of samples represent the direct output of the transmitter ANN. The trained transmitter can be viewed as a \emph{look-up table} which simply maps the input message to one of $M=64$ optimized blocks. Figure~\ref{fig:Tx_mapping} illustrates the 48 amplitude levels in each of these blocks. Interestingly, we see that the extremal levels $0$ and $\frac{\pi}{4}$ are the prevailing levels. It appears that the ANN tries to find a set of binary sequences optimized for end-to-end transmission. However, some intermediate values are also used. Unfortunately, it is not easy to say if this is intended by the deep learning optimization or an artefact. To bring more clarity, we visualize the constellation of modulation format by using state-of-the-art dimensionality reduction machine learning techniques such as t-Distributed Stochastic Neighbor Embedding (t-SNE)~\cite{tSNE}. Figure~\ref{fig:tSNE_constellation} shows the two-dimensional t-SNE representation of the un-filtered ANN outputs of Fig.~\ref{fig:Tx_mapping}. We can see that the 64 different waveforms are well-separated in the t-SNE space and can hence be discriminated well enough.} 

{Figure~\ref{fig:Spectrum} shows the spectrum of the real-valued electrical signal at the transmitter. Because of the low-pass filtering the spectral content is confined within 32\,GHz. The LPFs at both transmitter and receiver ensure that the signal bandwidth does not exceed the finite bandwidth of transmitter and receiver hardware. We can further observe that, as a result of the block-based transmission, the signal spectrum consists of strong harmonics at frequencies that are multiples of the symbol rate.  After DAC, modulation of the optical carrier, fiber propagation and direct detection by a PIN+TIA circuit, the samples of the distorted received waveforms are applied block-by-block as inputs to the receiver ANN for equalization.}

{\subsection{Comparison with Receiver-Only and Transmitter-Only ANN-Processing}}
{In contrast to systems with transmitter-only and receiver-only ANNs, the proposed end-to-end deep learning-based system enables joint optimization of the message-to-waveform mapping and equalization functions. To highlight the advantages of optimizing the transceiver in a single end-to-end process we compare---in simulation---our end-to-end design with three different system variations: (i) a system that deploys PAM2/PAM4 modulation and ANN equalization at the receiver; (ii) a system with ANN-based transmitter and a simple linear classifier at the receiver and (iii) a system with individually trained ANNs at both transmitter and receiver. In this section, we provide a detailed discussion on the implementation of each of these benchmark systems and relate their performance to the end-to-end deep learning approach. For a fair comparison all systems have a bit rate of 42\,Gb/s and 6\,bits of information are mapped to a block of 48 samples (including oversampling by a factor 4). All simulation parameters are as in Table~\ref{table1:Network parameters}. All hyper-parameters of the ANNs, such as hidden layers, activation functions, etc. as well as the other system and training parameters are identical to those used in the end-to-end learning system in Sec.~\ref{sec:simulation}.}

\begin{figure}[tb!]
\centering
\includegraphics[width=\columnwidth]{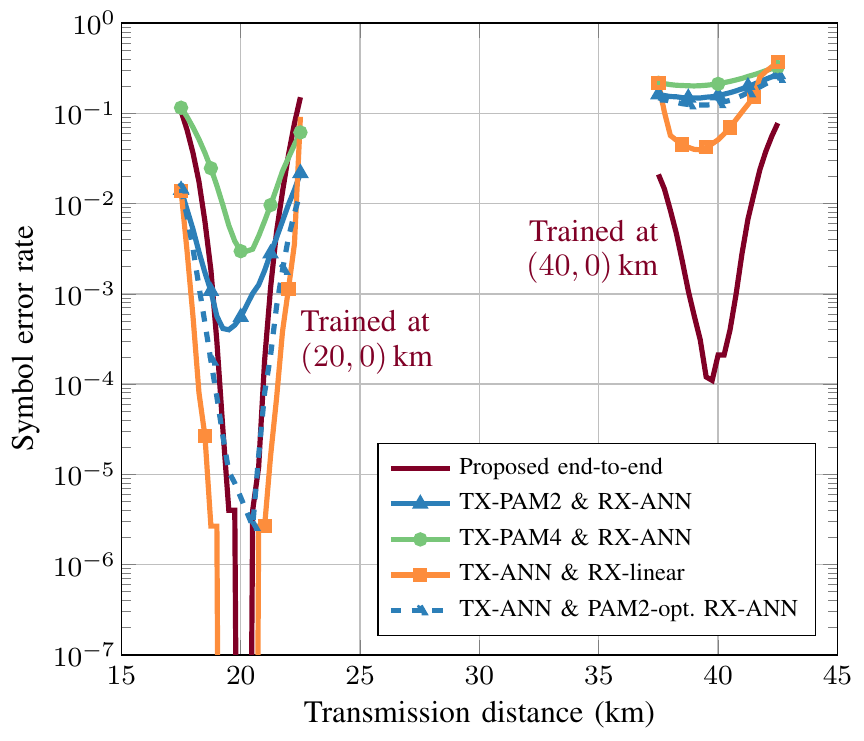}
\caption{{Symbol error rate as a function of transmission distance for (i) PAM2/PAM4 systems with ANN-based receiver, (ii) deep ANN-based transmitter and a \emph{multiclass-perceptron} receiver, (iii) ANN-based transmitter with ANN-based receiver optimized for PAM2 transmission and (iv) end-to-end deep ANN-based system. Training is performed at a fixed nominal distance of $20$\,km (left) or $40$\,km (right).}}
\label{fig:at_fixed_nominal_BLER}
\end{figure}

{\subsubsection{PAM transmitter \& ANN-based receiver} The PAM2 transmitter directly maps 6~bits into 6~PAM2 symbols ($\{0;\pi/4\}$). The PAM4 transmitter uses the best (6,3) linear code over GF(4)~\cite{Grassl} to map the 6~bits into 6~PAM4 symbols ($\{0;\pi/12;\pi/6;\pi/4\}$). The symbols are pulse-shaped by a raised-cosine (RC) filter with roll-off 0.25 and 2 samples per symbol. The waveform is further oversampled by a factor of 4 to ensure that a block of 48 samples is transmitted over the channel (as in the reference setup). The first element of the channel is the 32\,GHz LPF. The received block of distorted samples is fed to the ANN for equalization. Training of the receiver ANN is performed using the same framework as in Sec.~\ref{sec:simulation} by labeling the transmitted PAM sequences. Figure~\ref{fig:at_fixed_nominal_BLER} compares the symbol error rate performance of the described PAM2/PAM4 systems and the system trained in an end-to-end manner (curves ``TX-PAMx \& RX-ANN''). For training distances of 20\,km and 40\,km, the end-to-end ANN design significantly outperforms its PAM2 and PAM4 counterparts. In particular, at 20\,km the symbol error rate of the end-to-end system is below $10^{-6}$, while the PAM2 and PAM4 systems achieve $5.5\cdot10^{-4}$ and  $2.9\cdot10^{-3}$, respectively. At distances beyond 40\,km, the PAM-based systems with receiver-only ANN cannot achieve symbol error rates below $0.1$.}

\begin{figure}[tb!]
\centering
\includegraphics{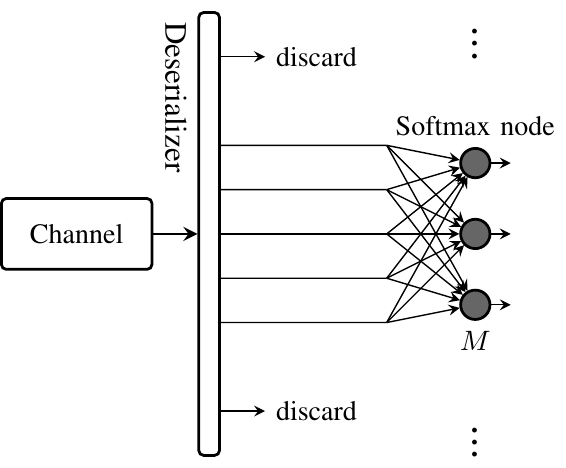}
\caption{{Schematic of the multiclass-perceptron used as receiver when having a deep ANN at the transmitter only.}}
\label{fig:perceptron}
\end{figure}

{\subsubsection{ANN-based transmitter \& linear receiver} In order to implement a system where the main ANN processing complexity is based at the transmitter, we employ the same ANN-based transmitter as in Fig.~\ref{fig:Schematic}. At the receiver, we impose a simple linear classifier as shown in Fig.~\ref{fig:perceptron}. This receiver is a linear classifier with $M$ classes, a so-called \emph{multiclass-perceptron} and carries out the operation $\mathbf{y}=\mathop{\textrm{softmax}}(\mathbf{W}_{\text{R}}\mathbf{x}+\mathbf{b}_{\text{R}})$, with $\mathbf{W}_{\text{R}}\in\mathbb{R}^{n\times M}$ and $\mathbf{b}_{\text{R}}\in\mathbb{R}^M$. The decision is made by finding the largest element of $\mathbf{y}$, i.e., $\hat{m} = \arg\max_{m \in \{1,\ldots,64\}}y_m$. The receiver thus employs only a single fully-connected layer with \emph{softmax} activation to transform the block of $n=48$ received samples into a probability vector of size $M=64$ (i.e. the size of the input \emph{one-hot vector}, see Sec.~\ref{sec:system}-A,C). At the transmitter, we use the exact same structure as in our deep ANN-based end-to-end design. Both the transmitter ANN parameters and the receiver parameters $\mathbf{W}_{\text{R}}$ and $\mathbf{b}_{\text{R}}$ are optimized in an end-to-end learning process. Hence, such a system exclusively benefits from the ANN-based pre-distortion of the transmitted waveform and has a low-complexity receiver. Figure~\ref{fig:at_fixed_nominal_BLER} also shows the performance of this system trained at distances 20\,km and 40\,km (``TX-ANN \& RX-linear''). The system trained at 20\,km achieves symbol error rate performance close to our deep learning-based end-to-end design. Moreover, we can see that it exhibits slightly better robustness to distance variations. This may be accounted to the absence of a deep ANN at the receiver, whose parameters during training are optimized specifically at the nominal distance and thus hinder the tolerance to distance changes. However, when the training is performed at 40\,km, this system exhibits a significantly inferior performance compared to the proposed end-to-end deep learning-based design.}

{\subsubsection{ANN-based transmitter \& ANN-based receiver, separately trained} Our final benchmark system deploys deep ANNs at both transmitter and receiver, which, in this case, are trained individually as opposed to performing a joint end-to-end optimization. For this comparison we fix the receiver ANN, whose parameters were previously optimized for PAM2 transmission, and aim to optimize only the transmitter ANN to match this given receiver in the best possible way. Training is carried out in the same end-to-end manner as detailed in Sec.~\ref{sec:simulation}, however, we keep the receiver ANN parameters fixed. Figure~\ref{fig:at_fixed_nominal_BLER} shows the symbol error rate performance of such a system (``TX-ANN \& PAM2-opt. RX-ANN''). For training at the nominal distance of 20\,km, this system design achieves a symbol error rate of $2.67\cdot10^{-6}$. Interestingly, one can clearly observe the benefits of the ANN-based waveform pre-distortion, which significantly lowers the error rate compared to the PAM2 system with receiver-only ANN. For systems trained at 40\,km however, the individually trained transmitter and receiver ANNs cannot outperform our proposed, jointly trained, end-to-end system.}

\begin{figure}[tb!]
\centering
\includegraphics[width=\columnwidth]{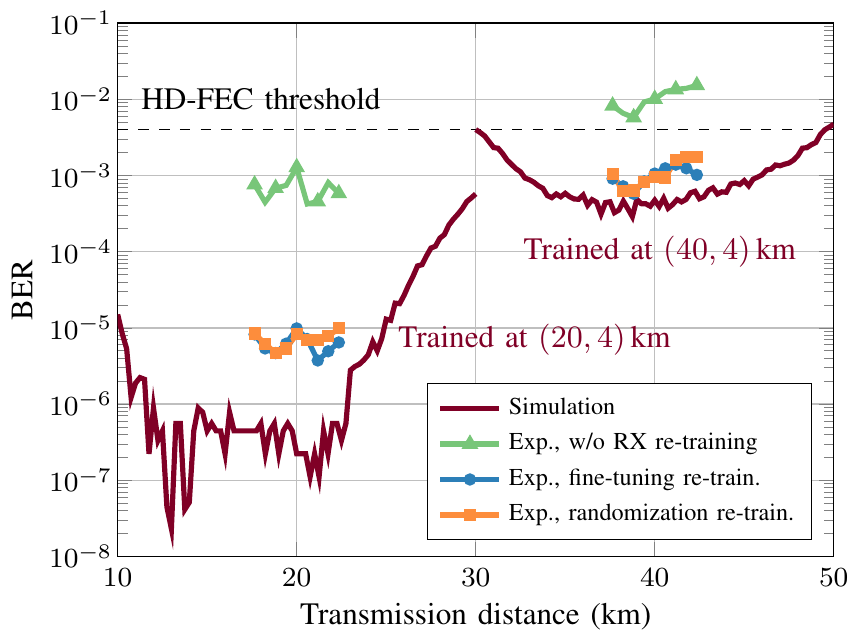}
\caption{Comparison of the experimental BER performance for systems trained at $(20,4)$\,km and $(40,4)$\,km (i) without re-training of the receiver ANN, (ii) re-training the receiver ANN by fine-tuning, (iii) training the receiver ANN by randomization. }
\label{fig:Discussion_Experiment_20_40}
\end{figure}

{\subsection{Further Details on the Experimental Validation}}
{As explained in Sec.~\ref{sec:experiment}, after propagation of the optimized waveforms during the experiment, the receiver ANN was fine-tuned (re-trained) to account for the discrepancies between the channel model used for training and the real experimental set-up. Re-training can be carried out in two different ways: In the first approach, denoted ``fine-tuning'', we initialize the receiver ANN parameters with the values previously obtained in simulation and then carry out re-training using the labeled experimental samples. In the second approach, denoted ``randomization'', we initialize the receiver ANN parameters with randomly initialized parameters sampled from a truncated normal distribution before re-training. Figure~\ref{fig:Discussion_Experiment_20_40} shows the experimental BER curves at 20 and 40\,km for the two re-training approaches and compares them with the raw experimental results, obtained by applying the initial Rx ANN acquired from the simulation 'as is' without any fine-tuning. We can observe that accounting for the difference between the real experimental environment and the assumed channel model by re-training improves performance at both distances. Moreover, expectedly, we confirm that both re-training solutions converge to approximately the same BER values at all examined distances. Although we kept the number of training iterations for the two approaches equal, initializing the ANN parameters with pre-trained values had the advantage of requiring less iterations to converge for most of the presented values.
It is also worth noting that the BER performance of the system without any re-training is well below the HD-FEC threshold around 20\,km, achieving a minimum value of $4.2\cdot10^{-4}$ at 20.59\,km. More accurate and detailed channel models used during training will likely further reduce this BER.}  

It is important to point out that for the experimental evaluation of ANN-based transmission schemes and hence in the framework of our work, the guidelines given in~\cite{Eriksson} need to be meticulously followed to avoid learning representations of a sequence (e.g., PRBS) used in the experiment and hence biasing the error rates towards too low values. In our work, during the offline training, we continuously generate new random input messages using a random number generator with a long sequence (e.g., Mersenne twister). In the experimental validation, we generated a long random sequence (\emph{not} a PRBS, as suggest in~\cite{Eriksson})  which  is processed by the transmitter ANN to generate a waveform, loaded (after filtering and resampling) into the DAC, and transmitted multiple times, to capture different noise realizations. For re-training the receiver ANN, \emph{mini-batches} are formed by picking randomly received blocks from a subset of the database of experimental traces (combining multiple measurements). 
Finally, in order to obtain the results presented throughout the manuscript, we use the trained and stored models to perform testing on a disjoint subset of the database of experimental traces, having no overlap with the subset used for training. This procedure ensures that the presented experimental results are achieved with independent data. Finally note that, due to the long memory of the fiber, it is not possible to capture the interference effects of all possible sequences of symbol preceding and succeeding and following the symbol under consideration in the experiment. Hence, it is possible that the results after re-training under-estimate the true error rate as the re-trained ANN may learn to adapt to the interference pattern of the sequence. Hence, the performance of all such ANN-based (re-trained) receivers can be considered to be a lower bound on the true system performance. Closely studying the effects of re-training based on repeated sequences is part of our ongoing work.

\section{Conclusion}\label{sec:conclusion}
For the first time, we studied and experimentally verified the  end-to-end deep learning design of optical communication systems. Our work highlights the great potential of ANN-based transceivers for future implementation of IM/DD optical communication systems tailored to the nonlinear properties of such a channel. We experimentally show that by designing the IM/DD system as a complete end-to-end deep neural network, we can transmit 42\,Gb/s beyond 40\,km with BERs below the 6.7\% HD-FEC threshold. The proposed system outperforms IM/DD solutions based on PAM2/PAM4 modulation and conventional receiver equalization for a range of transmission distances. Furthermore, we proposed and showed in simulations a novel training method that yields transceivers robust to distance variations that offer a significant level of flexibility. 
Our study is the first attempt towards the implementation of end-to-end deep learning for optimizing neural network based optical communication systems. As a proof of concept, we concentrated on IM/DD systems. We would like to point out that the method is general and can be extended to other, eventually more complex models and systems.

\section*{Acknowledgments}

The authors would like to thank Dr. Jakob Hoydis and Sebastian Cammerer for many inspiring discussions on the application of deep learning for communication systems.

The work of B. Karanov was carried out under the EU Marie Sk{\l}odowska-Curie project COIN (676448/H2020-MSCA-ITN-2015). The work of F. Thouin was carried out during an internship at Nokia Bell Labs supported by the German Academic Exchange Council under a DAAD-RISE Professional scholarship. The work of L. Schmalen was supported by the German Ministry of Education and Research (BMBF) in the scope of the CELTIC+ project SENDATE-TANDEM.

\ifCLASSOPTIONcaptionsoff
  \newpage
\fi

\end{document}